\documentclass[12pt]{article}
\usepackage{amsmath}
\usepackage{graphicx}
\usepackage{enumerate}
\usepackage{natbib}
\usepackage{url} 
\usepackage{bm}
\usepackage{float} 
\usepackage{subfigure}
\usepackage{algorithm}
\usepackage{algpseudocode}
\usepackage{xcolor}
\usepackage{hyperref}
\usepackage{amsthm}
\usepackage{amsfonts}
\usepackage{amsmath, bm, amssymb, booktabs}
\usepackage{tikz}
\usetikzlibrary{shapes,arrows}
\usepackage{dsfont}
\usepackage{titling}
\usepackage{bbm}
\usepackage{enumitem}
\usepackage{diagbox}
\usepackage{comment}
\usepackage{pgfplots}
\usepackage{tikz}
\usetikzlibrary{shapes,backgrounds}
\usepackage{booktabs}
\usepackage{tabularx}
\usepackage{ltablex} 
\usepackage{makecell} 
\usepackage{tikz}
\usetikzlibrary{positioning}
\usepackage{setspace}
\usepackage{hyperref}
\usepackage{xr}
\usepackage{bbm}
\usepackage{authblk}

\newcommand{\blind}{0}

\newtheorem{theorem}{\textbf Theorem}
\newtheorem{proposition}[theorem]{\textbf Proposition}

\DeclareMathOperator{\diag}{diag}

\begin{document}

\def\spacingset#1{\renewcommand{\baselinestretch}%
{#1}\small\normalsize} \spacingset{1}

\def\blind{1}
\if1\blind
{
  \title{\bf Generalized Heterogeneous Functional Model with Applications to Large-scale Mobile Health Data}
  \author[1]{Xiaojing Sun}
\author[2]{Bingxin Zhao}
\author[1]{Fei Xue}
\affil[1]{Department of Statistics, Purdue University}
\affil[2]{Department of Statistics and Data Science, University of Pennsylvania}
  \date{}
  \maketitle
} \fi

\if0\blind
{
  \bigskip
  \bigskip
  \bigskip
  \begin{center}
    {\LARGE\bf Title}
\end{center}
  \medskip
} \fi

\bigskip
\begin{abstract}

Physical activity is crucial for human health. With the increasing availability of large-scale mobile health data, strong associations have been found between physical activity and various diseases.
However, accurately capturing this complex relationship is challenging,  possibly because it varies across different subgroups of subjects,
especially in large-scale datasets.
To fill this gap, we propose a generalized heterogeneous functional method which simultaneously estimates functional effects and identifies subgroups within the generalized functional regression framework.
The proposed method captures subgroup-specific functional 
relationships between physical activity and diseases, providing a more nuanced understanding of these associations.
Additionally, we introduce a pre-clustering method that enhances computational efficiency for large-scale data 
through a finer partition of subjects compared to true subgroups.
In the real data application, we examine the impact of physical activity on the risk of mental disorders 
and Parkinson's disease using the UK Biobank dataset, which includes over 79,000 participants.
Our proposed method outperforms existing methods in future-day prediction accuracy,
identifying four  subgroups for mental disorder outcomes and three subgroups for Parkinson's disease diagnosis, with
detailed scientific interpretations  for each subgroup.
We also demonstrate theoretical consistency of our methods.
Supplementary materials are available online.
Codes implementing the proposed method are available at: \href{https://github.com/xiaojing777/GHFM}{https://github.com/xiaojing777/GHFM}.

\end{abstract}

\noindent%
\vfill

\newpage
\spacingset{1.9} 

{\color{black}
\section{Introduction}

{
Physical  activity plays a pivotal role in human health and is recognized as a factor that is highly related to the risk of both morbidity and mortality \citep{mahindru2023role, ekelund2019dose}. 
Modern advancements in technology, combined with the ubiquity of smartphones and wearable devices, have made access to physical activity data increasingly straightforward. 
In addition, large-scale mobile health datasets have become much more accessible over the past two decades \citep{althoff2017large}, such as the UK Biobank (UKB) dataset, which includes data from over 500,000 individuals \citep{sudlow2015uk}.
Many researchers study the relationship between physical activity and various diseases by analyzing large-scale datasets \citep{barker2019physical, rowlands2021association}. 
However, most of these studies focus on analyzing effects of physical activity on diseases that are assumed to be the same for all individuals, which might not be adequate for  large-scale datasets.

The heterogeneity in physical activity's effects across individuals could increase complexity 
of relationship between physical activity and diseases \citep{pearce2022association, rosmalen2012revealing}, and  this heterogeneity may arise from a range of complex factors. 
For example, differences in circadian rhythms may contribute to the effect heterogeneity:
the effect of physical activity on mortality for
individuals who go to bed early and get sufficient sleep 
could be different from that of
people who stay up late \citep{huang2022sleep}.
Exercise habits may also contribute to this variation: 
exercise beginners
often have more pronounced effects on health than people with long-established exercise routines \citep{garzon2017benefits}.
Although some studies  have investigated heterogeneity of physical activity \citep{albalak2023setting, shim2023wearable},
there are few studies considering heterogeneous effects of physical activity on diseases.


Moreover, existing studies focus mainly on summary statistics of physical activity, such as the minutes spent in moderate-to-vigorous physical activity \citep{rowlands2021association}.
In fact, physical activity is continuously varying across time, leading to a possibly time-dependent effect of physical activity on diseases. For instance, engaging in physical activity at midnight versus in the morning may lead to significantly different, or even opposite, effects on disease risk \citep{roig2016time}. 
To address this issue, 
functional data analysis (FDA) is a suitable method for capturing the time-dependent effects of physical activity \citep{ghosal2022scalar}. Some researchers have developed methods to account for heterogeneous functional effects \citep{yao2011functional, sun2022subgroup, zhang2022subgroup}. However, these approaches have limitations when applied to large-scale mobile health data. For example, the functional mixture model by \cite{yao2011functional} requires the number of subgroups to be specified in advance, which is often impractical. Additionally, the method proposed by \cite{zhang2022subgroup} only considers individualized intercepts rather than coefficient functions, making it inadequate for capturing the heterogeneous relationships between covariates and responses in FDA.

 }

In this paper, we  propose a generalized heterogeneous functional method and a pre-clustering procedure for large-scale data to investigate
the heterogeneous functional effects of physical activity on
diseases.
Specifically, in the generalized heterogeneous functional method, we propose to adopt subject-specific coefficient functions, rather than a single homogeneous coefficient function, to capture the varying effects of physical activity on diseases across individuals.  
We identify subgroups of subjects through 
fusing similar coefficient functions together based on a pairwise fusion penalty \citep{ma2017concave}, where the coefficient function in each subgroup is assumed to be homogeneous.
Although subject-specific effects account for individual heterogeneity, they introduce a large number of parameters for large samples, particularly in the FDA framework, which results in high computational costs.

To address this, we novelly propose a pre-clustering method for large-scale datasets, which reduces the number of parameters through 
grouping subjects into pre-clustering groups with most fitted coefficients, 
leading to a finer partition of true subgroups. 
We apply the proposed methods to analyze the functional effects of physical activity on mental disorders using neuroticism scores \citep{ormel2013neuroticism} and on Parkinson's disease within the UKB dataset.

The novelty and  advantages of the proposed method are as follows. 
First, it effectively handles large mobile health datasets. 
 In contrast to homogeneous methods, our proposed method allows and can identify subgroups in large datasets to account for potential time-varying heterogeneous effects of physical activity on diseases without requiring a predetermined true number of subgroups, providing a more accurate
and comprehensive understanding of their underlying relationship. 
Second, our proposed pre-clustering technique reduces computation cost when dealing with large-scale data. Our theoretical analysis shows that
the proposed pre-clustering groups provide a finer partition of the true subgroup identification, and it does not break the true subgroup structure.
Third, our method can handle not only continuous responses, such as neuroticism scores, but also generalized outcomes from exponential family distributions, which is important for real-world applications (e.g., binary Parkinson's disease diagnosis outcomes in the UKB study).

Fourth, in real data application, our method outperforms existing approaches in terms of future-day prediction accuracy.
The proposed method identifies three subgroups in a dataset of 80,692 subjects with Parkinson's disease diagnoses as outcomes, and finds  four subgroups in a dataset of 79,246 participants with neuroticism scores as outcomes.
Among the three subgroups for Parkinson's disease, one has
higher percentage of subjects diagnosed with Parkinson's disease, facilitating the detection of Parkinson's cases.
The analyses based on our identified subgroups are
consistent with 
existing studies \citep{chen2005physical, speelman2011might, bhidayasiri2018getting, beltagy2018night}. 
We also find that the optimal timing for physical activity to reduce disease risk varies across subgroups, offering practical recommendations for individualized interventions.

The remainder of this paper is organized as follows. In Section \ref{section: data}, we describe physical activity measurement and disease outcomes in the
UKB dataset. In Section \ref{section: method}, we propose the generalized heterogeneous functional method and the pre-clustering procedure. 
In section \ref{section: theory}, we illustrate theoretical results. Section \ref{section: simulation} provides numerical studies through simulations. In Section \ref{section: application}, we apply the proposed method to the UKB dataset.  
}

{\color{black}
\section{UKB Data}
\label{section: data}

Our work is motivated by the UKB dataset (\href{https://biobank.ctsu.ox.ac.uk/crystal/}{https://biobank.ctsu.ox.ac.uk/crystal/}), which is an extensive, publicly accessible, population-based, prospective cohort study that enrolled 502,490 participants between the ages of 40-69 during 2006-2010 \citep{sudlow2015uk}. 
In this paper, our goal is to explore functional relationship between physical activity and disease outcomes, 
while considering heterogeneity among participants. In this section, we provide an overview and pre-processing procedures for  
the physical activity data and disease outcomes in UKB.


\subsection{Physical Activity Data}
In this study, we utilize accelerometry data from the UKB to evaluate and analyze physical activity. Accelerometry is widely regarded as the gold standard for objectively measuring physical activity in population-based research \citep{strath2013guide}. Between February 2013 and December 2015, a total of 103,720 participants wore an Axivity AX3 wrist-mounted triaxial accelerometer for one week. This device continuously captured triaxial acceleration data at a frequency of 100 Hz, with a dynamic range of $\pm$8 mg (milli-g) \citep{doherty2017large}, as recorded in data field 90001 of the UKB.

The raw acceleration data were processed into five-second-epoch intervals, which are effective for differentiating between sedentary, light, moderate, and vigorous activities \citep{bai2016activity}. The initial step in this process is to calculate the Euclidean norm of  acceleration across the x, y, and z axes, which combines the acceleration into a single measure \citep{sabia2014association}. 
Then gravitational and noise effects were removed \citep{da2014physical, van2011estimation}. Finally, to assess the overall level and distribution of physical activity intensity, the data are aggregated into five-second intervals.
\citep{hammerla2013preserving}.

For participants with valid accelerometer data, we calculate average activity levels on an hourly basis 
using the five-second epoch acceleration data. This calculation produces a vector with 24 elements for each participant on each day, where each element corresponds to the average activity level for that specific hour, ranging from 00:00 (first element) to 23:00 (last element).
{
Figure \ref{fig: Four subjects PA plot } illustrates hourly
 activity levels of four individuals with distinct activity patterns. For example, subject 1 sleeps well at night, while subject 3 may have some sleep issues with high activity levels at night and naps during daytime.}
Previous research has demonstrated that analyzing hourly variations in activity over a 24-hour period using accelerometers provides valuable insights into the general adult population \citep{wennman2019gender, doherty2017large}.

We mainly focus on
the middle five days in the week of data collection
since the 24-hour cycles in first and last days of the week are incomplete.
We remove 7,122 subjects deemed of suboptimal quality by the UKB team, 
and further exclude 2,892 subjects 
whose wearing time is less than $70\%$ for at least one day among the middle five days in the week
\citep{leroux2021quantifying}. 
Our final sample size of physical activity data is  93,670.

\begin{figure}[h]
    \centering
    \includegraphics[width=0.6\textwidth]{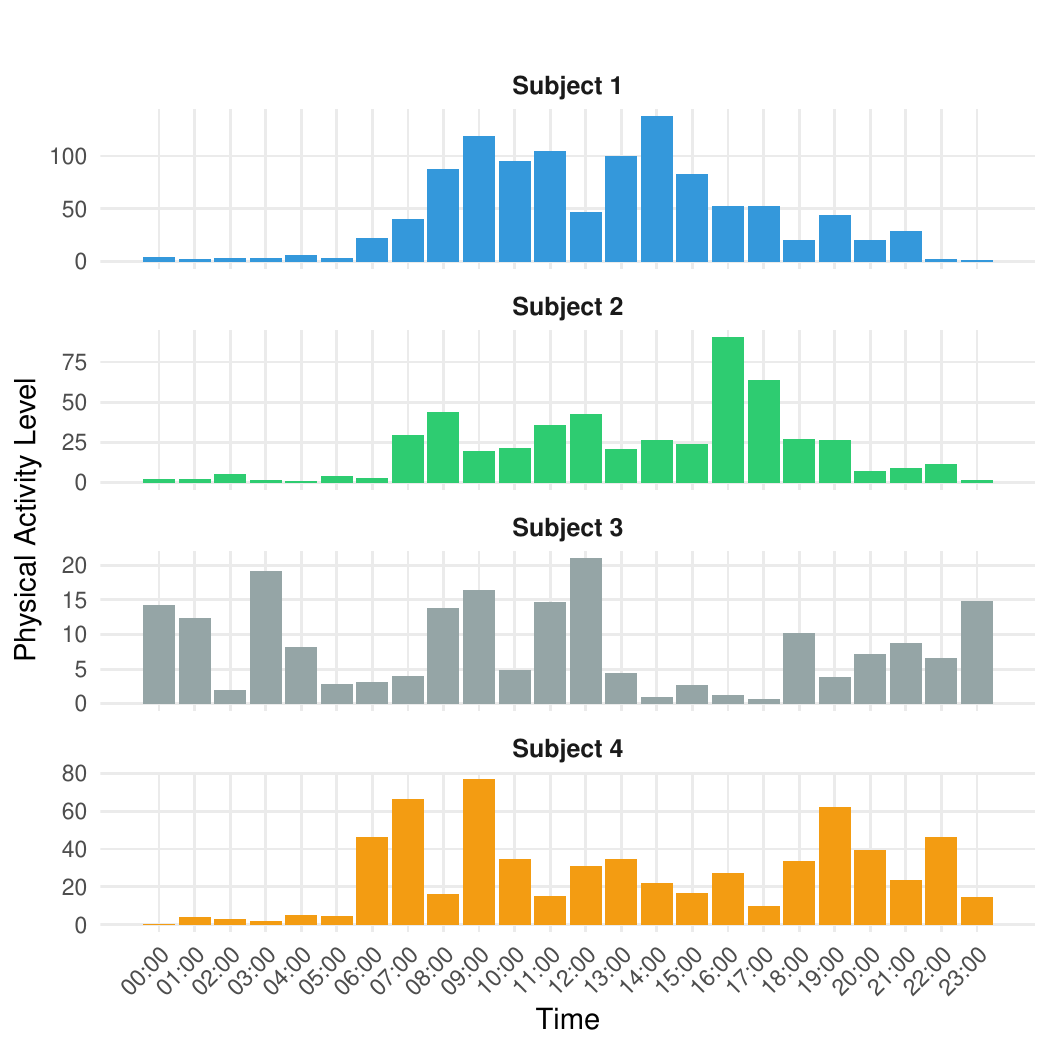}
    \caption{Illustration of the hourly average accelerometer activity level of four subjects.
    X-axis represents hourly time from 0:00 to 23:00.}
    \label{fig: Four subjects PA plot }  
\end{figure}


\subsection{Disease Data}

{In this paper, we consider neuroticism scores and Parkinson's disease as outcomes.}
We obtain diagnoses of Parkinson's disease for 446,829 subjects from ICD-10, specifically referencing field ID 41270. 
Among the 446,829 subjects, 
there are 4279 cases of Parkinson's disease, and
80,692 subjects with both physical activity and Parkinson's disease outcomes.

We consider the neuroticism score in the field 20127 of the UKB study as an indicator for mental diseases.
 This score of neuroticism is the sum of the number of ``Yes'' answers across twelve questions from different neurotic behaviour domains \citep{smith2013prevalence}, 
 where a higher value indicates unhealthier mental status.
 Examples of such questions include ``Do you ever feel `just miserable' for no reason?'' and ``Are you an irritable person?'' The complete list of questions is 
 at  \href{https://biobank.ctsu.ox.ac.uk/crystal/field.cgi?id=20127}{https://biobank.ctsu.ox.ac.uk/crystal/field.cgi?id=20127}.
 The mean neuroticism score is 4.119, with a standard deviation of 3.269.
There are 401,465 subjects with available neuroticism scores, among which
 79,252 subjects have both valid physical activity data and neuroticism score data.



}


{\color{black}
\section{Methodology}
\label{section: method}

In this section, we propose a generalized heterogeneous functional method (GHFM) to estimate  subgroup identification and coefficient functions simultaneously, and consider Gaussian and binary outcomes as specific examples. Moreover, we propose a pre-clustering method which improves computation efficiency for large sample while maintaining the true subgroup structure.
  The algorithm implementation of the proposed methods are provided in the   Supplementary Materials.

\subsection{Generalized Heterogeneous Functional Method (GHFM)}

In this subsection, we develop a generalized subject-wise scalar-on-function model to
study inherent heterogeneity among subjects. 
Let $X_{i1}(t), \dots , X_{ip}(t)$ be $p$ functional covariates and $Y_i$ be a scalar response of the $i$-th subject for $t\in[0,T]$ and $i=1,\dots, n$.
Suppose that the $p$ functional covariates are observed at $m$ discrete time grids $t_1, t_2, \ldots, t_{m}$,
and that the response $Y_i$ follows an exponential family distribution with density 
$f_{Y_i}(y_i)=\exp \{(y_i \theta_i-b(\theta_i)) / a(\psi)-c(y_i, \psi)\}, i=1,2,\dots,n$, where $\theta_i$ is a canonical parameter, $\psi$ is a dispersion parameter, and $a(\cdot), b(\cdot), c(\cdot,\cdot)$ are known functions depending on specific distributions.

Here we allow  the parameter $\theta_i$ to be subject-specific to account for heterogeneity among subjects. Moreover, we propose a generalized heterogeneous functional regression model
\setlength\abovedisplayskip{3pt}
\setlength\belowdisplayskip{4pt}
$$\eta(\mu_i)    = \alpha + \sum_{j=1}^p \int_{0}^{T} X_{ij}(t)\beta_{ij}(t) dt \  \text{ for } \  i=1,2,\dots, n,$$
where $\eta$ is a link function, $\mu_i = E(Y_i) = b'(\theta_i) $, $\alpha$ is the intercept term, and $\beta_{i1}(t), \dots, \beta_{ip}(t) $ are $p$ individualized  coefficient functions of the $i$-th subject defined on interval $[0,T]$. 
Rather than assuming a common coefficient function for each covariate in the traditional homogeneous scalar-on-function model, we consider subject-wise functional effects $\beta_{ij}(t)$ in the proposed model for $j=1,\dots, p$.

These subject-wise coefficient functions may exhibit certain subgroup structure of the entire population, which could be covariate-specific. 
Specifically, 
let $\mathcal{G}_j = \{\mathcal{G}_{j,1},\dots, \mathcal{G}_{j, \mathcal{K}_j} \}$ be a subgrouping partition of $\{1,\dots,n\}$
formed by coefficients $\beta_{1j}(t), \dots, \beta_{nj}(t)$ of the $j$-th covariate, where $\mathcal{K}_j$ is the number of distinct subgroups.
That is, for any $i, i' \in \{1,\dots,n\}$, we have $i, i' \in \mathcal{G}_{j,k}$ for some $k\in\{1,\dots, \mathcal{K}_j\}$ if and only if  $\beta_{ij}(t) = \beta_{i'j}(t)$.
Then the effect in each subgroup is homogeneous, while effects in different subgroups are distinct. This subgroup structure may vary across covariates.



To identify the subgroup structure and estimate  coefficient functions, we propose a pairwise grouping loss function
\begin{equation}
\label{loss func: general case}
\small
	Q_n(\alpha, \beta) = -\frac{1}{n}l(\alpha, \beta, \psi ; \mathbf{y}) + \phi \sum_{j=1}^p\sum_{i=1}^n  \int_0^{T} \left[\frac{d^2 \beta_{ij}(t)}{dt^2}\right]^2 dt + \lambda\sum_{j=1}^p \sum_{i\neq i'} \int_0^T |\beta_{ij}(t)-\beta_{i'j}(t)|dt,
\end{equation}
where 
{
\setlength\abovedisplayskip{2pt}
\setlength\belowdisplayskip{4pt}
$$l(\alpha, \beta, \psi ; \mathbf{y})=\sum_{i=1}^n l_i\left(\theta_i, \psi ; y_i\right)=\sum_{i=1}^n\left\{\left[y_i \theta_i(\alpha, \beta_i)-b\left(\theta_i(\alpha, \beta_i)\right)\right] / a(\psi)-c_i\left(y_i, \psi\right)\right\}$$
}
is the log likelihood function,
$\mathbf{y} = (Y_1,\dots,Y_n)^T, \beta_i = (\beta_{i1}(t), \dots, \beta_{ip}(t))^T$, $\beta = (\beta_{1}^T, \dots, \beta_{n}^T)^T$, and  $\phi, \lambda$ are tuning parameters.

Here we adopt a roughness penalty (the second term in the loss function) to control the fluctuation of each coefficient function $\beta_{ij}(t)$ \citep{green1993nonparametric}, and a functional pairwise fusion penalty (the third term in the loss function) to fuse similar subject-wise effects together.
Specifically, the fusion penalty encourages subjects to share the same coefficient function when their corresponding effects are similar. 
In this way, we can reduce the number of coefficients, identify subgroups, and borrow information across subjects to estimate the coefficient functions accurately.


For estimation of the coefficient functions,
we propose to approximate $\beta_{ij}(t)$
through a linear combination of B-spline basis functions  for $i=1,\dots,n$ and $j=1,\dots,p$. 
Specifically, we approximate $\beta_{ij}(t) $ as ${\beta}_{ij}(t) \approx {\bm{b}}_{ij}^T \bm{B}(t)$,
where ${\bm{b}}_{ij}$ is an $L$-dimensional coefficient vector, 
$\mathbf{B}(t)$ is an $L$-dimensional vector of B-spline basis functions of order $(d+1)$ with $(M+1)$ equally spaced knots on the time interval $[0,T]$, and $L=M+d $. 
Given this approximation, the loss function in Equation (\ref{loss func: general case}) becomes
 \begin{equation}
 \small
 \setlength\abovedisplayskip{3pt}
\setlength\belowdisplayskip{2pt}
 	\label{loss func: general case b}
 	D_n(\alpha, \bm{b}_1, \dots, \bm{b}_p) = -\frac{1}{n}\tilde{l}(\alpha, \bm{b}_1, \dots, \bm{b}_p, \psi ; \mathbf{y}) + \phi\sum_{j=1}^p \bm{b}_j^TR\bm{b}_j + \lambda \sum_{j=1}^p  \sum_{i\neq i'} \int_0^T | \bm{B}^T(E_i - E_{i'})^T \bm{b}_j |dt,
 \end{equation}
 where $\bm{b}_j = ({\bm{b}}_{1j}^T,\dots, {\bm{b}}_{nj}^T)^T $ is an $(nL)$-dimensional column vector, $\tilde{l}(\alpha, \bm{b}_1, \dots, \bm{b}_p, \psi ; \mathbf{y})$ is the value of $l(\alpha, \beta, \psi ; \mathbf{y})$ with each $\beta_{ij}$ replaced by ${\bm{b}}_{ij}^T \bm{B}(t)$,
 \begin{equation}
\small
R = \diag\left\{\int _0^{T}  \frac{d^2 \mathbf{B}(t)}{dt^2} \frac{d^2 \mathbf{B}^T(t)}{dt^2} dt, \dots, \int _0^{T}  \frac{d^2 \mathbf{B}(t)}{dt^2} \frac{d^2 \mathbf{B}^T(t)}{dt^2} dt\right\}
\label{Matrix: R}
\end{equation}
is a block diagonal matrix with $n$ $L\times L$ blocks on the diagonal, and 
\begin{equation}
\setlength\abovedisplayskip{3pt}
\setlength\belowdisplayskip{4pt}
\label{def_E_i}
    E_i=(\bm{0}_{L\times L}, \dots, \bm{I}_{L\times L}, \dots, \bm{0}_{L\times L})^T
\end{equation}
is an $nL\times L$ matrix consisting of $n$  $L\times L$ blocks with the $i$-th block being an identity matrix of size $L$ and the other blocks being zero matrices.

We obtain the proposed estimator for coefficient functions by $\hat{\beta}_{ij}(t) = \hat{\bm{b}}_{ij}^T \bm{B}(t)$, where $\hat{\bm{b}}_{ij}$ is the $\bm{b}_{ij}$-coordinate of the minimizer of the loss function in Equation (\ref{loss func: general case b}).
Subject $i$ and Subject $i'$ are estimated to belong to the same subgroup for the $j$-th covariate if and only if $\hat{\beta}_{ij}(t) = \hat{\beta}_{i'j}(t)$ for $ i, i'=1,\dots,n$ and $j=1,\dots,p$.

\subsection{Heterogeneous Functional Gaussian Method (HFGM)}

In this subsection, we consider the proposed method with Gaussian outcomes,  and refer to it as the heterogeneous functional Gaussian method (HFGM). Specifically, we assume $Y_i \sim  N(\mu_i, \sigma^2)$ for $ i=1,2,\dots,n$, {
implying that $\theta_i = \mu_i, b(\theta_i) = \mu_i^2/2, \eta(\mu_i) = \mu_i, a(\psi) = \sigma^2,$ and $c(y_i, \psi)=\left(y_i^2 / \sigma^2+\log 2 \pi \sigma^2\right) / 2$. 
}
For this Gaussian outcome, the proposed generalized heterogeneous functional regression model becomes
\begin{equation}
\setlength\abovedisplayskip{3pt}
\setlength\belowdisplayskip{4pt}
\label{model: HFGM}
Y_i = \alpha + \sum_{j=1}^p \int_{0}^{T} X_{ij}(t)\beta_{ij}(t) \, dt + \epsilon_i \ \text{ for } \  i=1,2,\dots,n,
\end{equation}
where $\epsilon_i \sim  N(0, \sigma^2).$ 
The corresponding Gaussian log likelihood is $l(\alpha, \beta, \psi ; \mathbf{y}) = \sum_{i=1}^n\{-(Y_i - \alpha - \sum_{j=1}^p \int_{0}^{T} X_{ij}(t)\beta_{ij}(t) dt)^2/2\sigma^2 - \log \left(2 \pi \sigma^2\right) / 2 \}$. 

Since maximizing $l(\alpha, \beta, \psi ; \mathbf{y})$ with respect to $\alpha$ and $\beta(t)$ is equivalent to minimizing the functional least squares term $ {l}^*(\alpha, \beta, \psi ; \mathbf{y}) = \sum_{i=1}^n(Y_i - \alpha - \sum_{j=1}^p \int_{0}^{T} X_{ij}(t)\beta_{ij}(t) dt)^2$, we propose the following loss function for the HFGM:
\begin{equation}\small
\setlength\abovedisplayskip{4pt}
\setlength\belowdisplayskip{4pt}
\label{loss func: gauss beta t}
Q_n^{G}(\alpha, \beta) = \frac{1}{n}{l}^*(\alpha, \beta, \psi ; \mathbf{y}) + \phi \sum_{j=1}^p \sum_{i=1}^n \int_0^{T} \left(\frac{d^2 \beta_{ij}(t)}{dt^2}\right)^2 dt + \lambda \sum_{j=1}^p \sum_{i\neq i'} \int_0^T |\beta_{ij}(t)-\beta_{i'j}(t)|dt.
\end{equation}
Using the B-spline basis functions, we can approximate this loss function by
{\small
\setlength\abovedisplayskip{3pt}
\setlength\belowdisplayskip{4pt}
\begin{align}
    D_n^G(\alpha, \bm{b}_1, \dots, \bm{b}_p) =& \frac{1}{n} \sum_{i=1}^n \left( Y_i - \alpha - \sum_{j=1}^p \bm{\gamma}_{ij}^T E_i^T \bm{b}_j  \right)^2 + \phi \sum_{j=1}^p \bm{b}_j^TR\bm{b}_j \notag\\
    &+ \lambda \sum_{j=1}^p \sum_{i\neq i'} \int_0^T | \bm{B}^T(E_i - E_{i'})^T \bm{b}_j |dt, \label{loss func: gaussian b}
\end{align}
}
where $\bm{\gamma}_{ij} = \int_0^T \bm{B}(t)X_{ij}(t)dt $ is an $L$-dimensional column vector. 

\subsection{Heterogeneous Functional Logistic Method (HFLM)}

In this subsection, we consider the proposed method with binary outcomes which are common in practical scenarios, particularly in medical health where diagnosed cases are denoted as "1" and non-diagnosed cases as "0". 
We refer to the proposed method for binary outcomes as the heterogeneous functional logistic  method (HFLM). 
Here we assume that the binary response $Y_i$ follows the Bernoulli distribution with probability $p_i$ for $i=1,2,\dots, n$. Then the heterogeneous functional logistic regression model is of the form
\begin{equation}
\setlength\abovedisplayskip{3pt}
\setlength\belowdisplayskip{4pt}
\label{model: HFLM}
    \ln [{p_i}/{(1-p_i)}] = \alpha + \sum_{j=1}^p \int_{0}^{T} X_{ij}(t)\beta_{ij}(t) dt,  i=1,\dots, n.
\end{equation}
The corresponding loss function for the HFLM is
\begin{equation}
\small
\setlength\abovedisplayskip{3pt}
\setlength\belowdisplayskip{4pt}
\label{loss func: logistic beta t}
Q_n^L(\alpha, \beta) = -\frac{1}{n}l(\alpha, \beta, \psi ; \mathbf{y}) + \phi \sum_{j=1}^p \sum_{i=1}^n  \int_0^{T} \left[\frac{d^2 \beta_{ij}(t)}{dt^2}\right]^2 dt + \lambda \sum_{j=1}^p \sum_{i\neq i'} \int_0^T |\beta_{ij}(t)-\beta_{i'j}(t)|dt,
\end{equation}
where
{\small
\begin{equation*}
	\setlength\abovedisplayskip{1pt}
	\setlength\belowdisplayskip{4pt}
	l(\alpha, \beta, \psi ; \mathbf{y}) = \sum_{i=1}^n \left\{ y_i\left[\alpha+ \sum_{j=1}^p\int\beta_{ij}(t)X_{ij}(t)dt\right] - \ln\left[1+\exp\left(\alpha+\sum_{j=1}^p\int\beta_{ij}(t)X_{ij}(t)dt\right) \right]  \right\}.
\end{equation*}}

\noindent Using the B-spline basis functions, we approximate this loss function by
\begin{equation}
\small
\setlength\abovedisplayskip{3pt}
\setlength\belowdisplayskip{4pt}
\label{loss func b: logis}
D_n^L(\alpha,\bm{b}_1, \dots, \bm{b}_p) = -\frac{1}{n}\tilde{l}(\alpha, \bm{b}_1, \dots, \bm{b}_p, \psi ; \mathbf{y}) + \phi \sum_{j=1}^p \bm{b}_j^TR\bm{b}_j + \lambda \sum_{j=1}^p  \sum_{i\neq i'} \int_0^T | \bm{B}^T(E_i - E_{i'})^T \bm{b}_j |dt,
\end{equation}
where
$\tilde{l}(\alpha, \bm{b}_1, \dots, \bm{b}_p, \psi ; \mathbf{y}) = \sum_{i=1}^n [ y_i (\alpha+ \sum_{j=1}^p \gamma _{ij}^TE_i^T\bm{b}_j) - \ln(1+\exp{(\alpha + \sum_{j=1}^p \gamma _{ij}^T E_i^T\bm{b}_j)} )] $. 

\subsection{Pre-clustering}

{
Since parameters in the proposed loss functions in Equations (\ref{loss func: general case}) and (\ref{loss func: general case b}) are subject-wise, the minimization of the proposed loss could be computationally challenging when the sample size is large. In fact, this kind of large-sample or large-scale data are common in practice.
For instance, the number of subjects in the UKB study with available physical activity data is  93,670. }


To solve this issue, we propose a pre-clustering method, which provides an initial clustering of all subjects.
A pre-clustering is a partition of all subjects into
$K$ potential groups such that, 
if the $i$-th subject and the $i'$-th subject are in the same potential group, then we have $\beta_{ij}(t) = \beta_{i'j}(t)$ for all $j=1,\dots, p$. That is, subjects in the same potential group belong to the same true subgroup for each covariate. 
We refer to these potential groups as pre-clustering groups. 
Note that
$K\ge\prod_{j=1}^p\mathcal{K}_j$.

Subjects in the same true subgroup might not be in the same pre-clustering group, which implies that the pre-clustering group structure is a finer partition of all subjects than the true subgroup structure.
Figure \ref{fig:Relationship between true subgroups and pre-clustering groups. } illustrates this through an example with two true subgroups and five pre-clustering groups, where the first true subgroup consists of three pre-clustering groups, and the second true subgroup consists of two pre-clustering groups.
Moreover, the pre-clustering group structure is not covariate-specific, that is, there is only one rather than $p$ pre-clustering group structures.

\begin{figure}[h]
	\centering
	\includegraphics[width=0.75\textwidth]{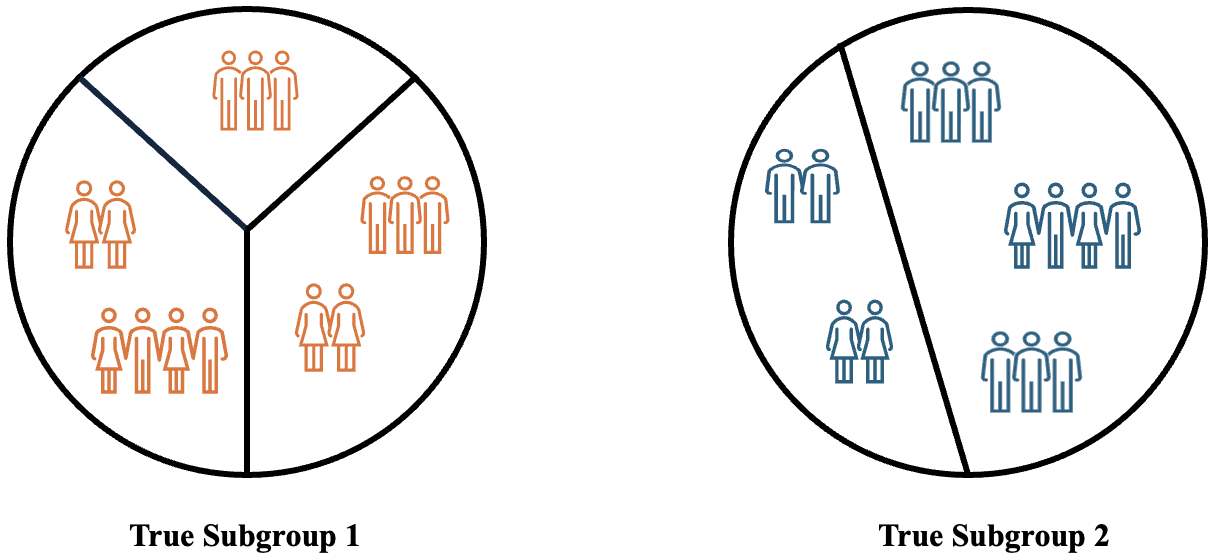}
	\caption{An example for relationship between true subgroups and pre-clustering groups when there are two true subgroups and five pre-clustering groups.}
	\label{fig:Relationship between true subgroups and pre-clustering groups. }
\end{figure}

{

We propose a pre-clustering loss function
\begin{equation}
\setlength\abovedisplayskip{1pt}
\setlength\belowdisplayskip{4pt}
\small
    L(\alpha, \beta_{(1)},\dots,\beta_{(K)}) = \sum_{i=1}^{n} \min_{k= 1,\dots,K} L_{i,k}(\alpha, \beta_{(k)}),
\label{loss function: pre-clustering}
\end{equation}
where $\alpha$ is the intercept term, $\beta_{(k)} = (\beta_{(k),1}(t), \dots, \beta_{(k),p}(t))^T$ with $\beta_{(k),j}(t)$ denoting the coefficient function  in the $k$-th pre-clustering group for the $j$-th covariate, and $L_{i,k}(\alpha, \beta_{(k)}) = -\left[y_i \theta_i(\alpha, \beta_{(k)})-b\left(\theta_i(\alpha, \beta_{(k)})\right)\right] / a(\psi)+c_i\left(y_i, \psi\right)$ is the negative log-likelihood of the $i$-th subject with parameters $\alpha$ and $\beta_{(k)}$ from the $k$-th pre-clustering group. 
Let 
\begin{equation}
\setlength\abovedisplayskip{3pt}
\setlength\belowdisplayskip{4pt}
\label{minimizer: pre-clustering}
(\tilde{\alpha}, \tilde{\beta}_{(1)},\dots,\tilde{\beta}_{(K)})  = 
\underset{\alpha \in \mathbb{R}, {\beta}_{(k),j}(t) \in S_{dM}, k\in\{1,\dots,K\}, j\in\{1,\dots,p\} }{\mathrm{argmin}}
L({\alpha}, {\beta}_{(1)},\dots,{\beta}_{(K)}),
\end{equation}
be the minimizer of the pre-clustering loss function, where $\mathcal{S}_{d M}$ denotes the space spanned by B-spline basis functions of degree $d+1$ and knots $M+1$.

Here we identify the pre-clustering group label of the $i$-th subject through finding out the smallest $L_{i,k}$ among all pre-clustering groups ($k=1,\dots, K$) since a smaller loss indicates better fitness between the subject and a pre-clustering group.
Following this rule, we can obtain pre-clustering groups $\tilde{\mathcal{G} }= \{ \tilde{\mathcal{G}}_{1},\dots, \tilde{\mathcal{G}}_{K} \}$
identified by the minimizer $(\tilde{\alpha}, \tilde{\beta}_{(1)},\dots,\tilde{\beta}_{(K)})$, 
{where $\tilde{\mathcal{G}}_{k}$ is the index set of subjects in the $k$-th pre-clustering group for $k=1,\dots, K$.}

Compared with the loss function in Equation (\ref{loss func: general case}), the above pre-clustering loss function reduces the number of coefficient functions from $n$ to $K$ for each covariate since the coefficient functions in Equation (\ref{loss function: pre-clustering}) are group-specific instead of subject-specific, which reduces computation cost.
In Theorem \ref{thm: pre-clustering} of Section \ref{section: theory} below, we will show that minimizing  the pre-clustering loss can lead to an estimated pre-clustering
group identification that does not break the true subgroup structure 
under some regularity conditions.

}

After the pre-clustering groups are estimated, we apply the proposed pairwise grouping loss function in Equation (\ref{loss func: general case}) to the identified pre-clustering groups, and 
minimize
the following loss function
\begin{equation}
\setlength\abovedisplayskip{3pt}
\setlength\belowdisplayskip{4pt}
\begin{split}
\label{loss func: after pre-clustering}
\small
Q_n(&\alpha,  \beta_{(1)},\dots,\beta_{(K)} ) = -\Bar{l}(\alpha,  \beta_{(1)},\dots,\beta_{(K)}, \psi ; \mathbf{y}) \\
&\quad + \phi \sum_{j=1}^p \sum_{k=1}^K  \int_0^{T} \left(\frac{d^2 \beta_{(k),j}(t)}{dt^2}\right)^2 dt  + \lambda \sum_{j=1}^p \sum_{k\neq k'} \int_0^T |\beta_{(k),j}(t)-\beta_{(k'),j}(t)|dt,
\end{split}
\end{equation}
where 
$\Bar{l}(\alpha,  \beta_{(1)},\dots,\beta_{(K)}, \psi ; \mathbf{y}) = \sum_{k=1}^K  \sum _{i\in \tilde{\mathcal{G}}_{k}} \left\{\left[y_i \theta_i(\alpha, \beta_{(k)})-b\left(\theta_i(\alpha, \beta_{(k)})\right)\right] / a(\psi)-c_i\left(y_i, \psi\right)\right\} $.
{
Subjects in the $k$-th pre-clustering group and in the $k'$-th pre-clustering group are estimated to belong to the same subgroup for the $j$-th covariate if and only if $\hat{\beta}_{(k),j}(t) = \hat{\beta}_{(k'),j}(t)$, where $\hat{\beta}_{(k),j}(t)$ is the $\beta_{(k),j}(t)$-coordinate of the minimizer of the loss in Equation (\ref{loss func: after pre-clustering}) for $ k, k'=1,\dots,K$ and $j=1,\dots,p$.
Consequently, $\hat{\beta}_{(k),j}(t)$ is the estimated coefficient function in this subgroup.
}

\section{Theory}
\label{section: theory}


\subsection{Consistency of GHFM}
In this subsection, 
we  demonstrate consistency of subgroup identification and parameter estimation of the proposed GHFM.

We  first introduce some notations and 
regularity conditions.
We let $\lambda_n=\lambda$ and  $\phi_n=\phi$ in this section to emphasize that the tuning parameters may change as sample size $n$ increases. 
We define the $L_2$ norm and infinity norm of a continuous function $h$ on the interval $[0, T]$ as 
$
\|h(t)\|_2 = \left(\int_0^T h^2(t) \, dt\right)^{1/2}
$
and 
$\|h (t)\|_{\infty}=\sup \{|h(t)|: t \in[0, T]\}$
respectively.
We use  $f(n) \gtrsim g(n)$ to denote that there exists a constant $C > 0$ such that $f(n) \geq C \cdot g(n)$ for sufficiently large $n$. 
We let $\lambda_{\min}(A)$ and $\lambda_{\max}(A)$ represent the minimum and maximum eigenvalues of matrix $A$, 
and $J_{n,j} = \diag\{\bm{\gamma}_{1j}\bm{\gamma}_{1j}^T, \dots, \bm{\gamma}_{nj}\bm{\gamma}_{nj}^T\}$.



(C1) We assume that $\|X_j(t)\|_2$ is almost surely  bounded above by a constant for $j=1,\dots,p$, and that $X_{1j}(t),\dots,X_{nj}(t)$ are $n$  independent and identically distributed (i.i.d.) realizations of the stochastic process $X_j(t)$.



(C2) Suppose that, for each $i\in \{1,\dots,n\}$ and $j\in \{1,\dots,p\}$, the coefficient function $\beta_{ij}(t)$ is in the Hölder space $C^{p^*, v}$ for some positive integer $p^*$ and $v \in[0,1]$ such that $p^*+v\le d$ .
That is, 
there is a constant $C_1>0$ such that
$\mid \beta_{ij}^{\left(p^*\right)}\left(u_1\right)-\beta_{ij}^{\left(p^*\right)}\left(u_2\right)\left|\leq C_1\right| u_1-\left.u_2\right|^v$ for any
$u_1, u_2 \in [0,T]$, where $\beta_{ij}^{\left(p^*\right)}\left(t\right)$ denotes the $p^*$-th derivative of $\beta_{ij}(t)$.


(C3) 
We assume that the number of time grids $M=o\left(n^{1 / \psi_1 }\right)$, $M \gtrsim n^{1 / \psi_2 }$, 
tuning parameters 
$\phi_n=o\left(n^{-1 / 2}\right)$ and $\lambda_n=o(n^{-3/2} M^{-1})$, where $\psi_2>\psi_1>2$.

{
(C4) 
We assume $\max \{\bm{\gamma}_{1j}^T\bm{\gamma}_{1j}, \dots, \bm{\gamma}_{nj}^T\bm{\gamma}_{nj}\} \gtrsim M^{-1/2}$ almost surely for $j\in\{1,\dots,p\}$.

}


(C5) We assume the  number of true subgroups 
$\mathcal{K}_j = O(1)$ and $\min_{k=1,\dots,\mathcal{K}_j} | \mathcal{G}_{j,k} | \gtrsim n$  for $j\in\{1,\dots,p\}$.

(C6) 
Suppose that $\min _{i \in \mathcal{G}_{j,k}, i' \in \mathcal{G}_{j, k'}, k \neq k^{\prime}}\left\|\beta_{ij}(t)-\beta_{i'j}(t)\right\|_2 > C_3 
$ for some constant $C_3 > 0$.

 Conditions (C1) and (C2) are the same as Hypothesis (H1) and (H3) of \cite{cardot2003spline} which are standard assumptions for non-parametric B-spline methods \citep{shen1998local, xue2010consistent, claeskens2009asymptotic}. Condition
(C1) ensures the identifiability of penalized functional linear model, that is, the existence
and uniqueness of the coefficient functions \citep{bosq2000linear}.
Condition (C3) specifies the exact requirements for the tuning parameters.
Condition (C4) is equivalent to requiring that $\lambda_{\max}(J_{n,j}) \gtrsim M^{-1/2}$, which is similar to Condition (C4) in \cite{lin2017locally} and Condition (A8) in \cite{zhou2013functional}.}
{\color{black}
Condition (C5) indicates that the number of true subgroups does not increase as the sample size goes to infinity, while the size of each subgroup expands with increasing samples. 
Condition (C6)  provides a lower bound of the minimum distance of coefficient functions in different subgroups,
which is crucial for identifying heterogeneous subgroups \citep{ma2017concave}.


The following Theorem \ref{thm: consistency of HFGM} shows that with probability tending to one, there exists a local solution of Equation (\ref{loss func: gauss beta t}) that converges to $\beta_{1j}(t), \ldots, \beta_{nj}(t)$, and the corresponding estimated subgroup identification $\{\hat{\mathcal{G}}_{j,1},\dots, \hat{\mathcal{G}}_{j, \hat{\mathcal{K}}_j} \}$ converges to true subgroup identification $\{\mathcal{G}_{j,1},\dots, \mathcal{G}_{j, \mathcal{K}_j} \}$ for $j\in \{1,\dots,p\}$.

\begin{theorem}
\label{thm: consistency of HFGM}
When Conditions (C1) to (C6) are satisfied, for the loss function of the HFGM in Equation (\ref{loss func: gauss beta t}),
with probability tending to one, there exists a local minimizer  $(\hat{\alpha}, \hat{{\beta}}) = \arg \min_{\alpha\in\mathbb{R}, \beta \in \mathcal{S}_{d M}} Q_{n}^G({\alpha, \beta})$  such that 
$ \| \hat{\beta}_{ij}(t) - \beta_{ij}(t) \|_\infty < \xi_n$, where $\{\xi_n\}$ is a sequence converges to 0,
and  
$\operatorname{Pr}\left( \{\hat{\mathcal{G}}_{j,1},\dots, \hat{\mathcal{G}}_{j, \hat{\mathcal{K}}_j} \} =  \{\mathcal{G}_{j,1},\dots, \mathcal{G}_{j, \mathcal{K}_j} \} \right) \rightarrow 1$ for $i\in\{1,\dots,n\}, j\in \{1,\dots,p\}$,
where $ \hat{{\beta}} = (\hat{\beta}_{11}(t),\dots,\hat{\beta}_{np}(t))^T$.
\end{theorem}

The theorem demonstrates the consistency of both coefficient function estimation and subgroup identification. 
In fact, the consistency of subgroup identification naturally follows from the consistency of coefficient function estimation,
since 
$\| \hat{\beta}_{ij}(t) - \hat{\beta}_{i'j}(t) \|_\infty \le 2\underset{k}{\max}
\| \hat{\beta}_{kj}(t) - \beta_{kj}(t) \|_\infty$ converges to zero when $\mathcal{S}_j(i) = \mathcal{S}_j(i')$, and
$\| \hat{\beta}_{ij}(t) - \hat{\beta}_{i'j}(t) \|_\infty \geq 
\underset{\mathcal{S}_j(i) \neq \mathcal{S}_j(i')}{\min}
\| \beta_{ij}(t) - \beta_{i'j}(t) \|_\infty - 
2\underset{k}{\max}
\| \hat{\beta}_{kj}(t) - \beta_{kj}(t) \|_\infty>0$ for large sample when
$\mathcal{S}_j(i) \neq \mathcal{S}_j(i')$,
where {
$\mathcal{S}_j(i)$}
denotes the true subgroup index of the $i$-th subject with respect to the $j$-th covariate.
Consistency of HFLM is illustrated in {the  Supplementary Materials}. 
Theoretical results for other generalized outcomes could be derived in a similar manner.

\subsection{Theory for Pre-clustering}

In this subsection, we 
provide a theoretical guarantee of our implementation of  
the proposed pre-clustering 
and establish properties of the pre-clustering group identification for Gaussian reponses.
The detailed steps of implementation of the proposed pre-clustering  are provided  in the Supplementary Materials.

In Proposition \ref{thm: alg converge (SSE)} below, 
we provide the convergence of 
the algorithm for pre-clustering in the Supplementary Materials, 
requiring the following additional regularity condition (C7) which assumes that the number of true subgroups is much smaller than that of pre-clustering groups. 

{

(C7) The number of pre-clustering groups $K = O(1)$, and it satisfies that $\prod_{j=1}^p \mathcal{K}_j<<K$ for $j\in\{1,\dots,p\}$.
}

\begin{proposition}
\label{thm: alg converge (SSE)}
When Conditions (C1), (C2), (C5), (C6), and (C7) are satisfied, we have
$\operatorname{SSE}^{(l+1)}
< \operatorname{SSE}^{(l)}
$ 
for $ l=0,1,2, \dots$, 
where $l$ denotes the number of iterations in the algorithm for pre-clustering, $\operatorname{SSE}^{(l)} = \sum_{i=1}^{n} (y_i - \hat{y}_i^{(l)})^2$, 
$\hat{y}_i^{(l)} = \tilde{\alpha}^{(l)} + \sum_{j=1}^p \int_0^T \tilde{\beta}_{(k)j}^{(l)}(t)X_{ij}(t) dt $, and $k\in \{1,\dots,K\}$ is 
the pre-clustering group index of the $i$-th subject at the $l$-th iteration. 
\end{proposition}
 The strict decrease in SSE in Proposition \ref{thm: alg converge (SSE)} implies the convergence of the {
\color{black}
algorithm. 
}

To demonstrate the consistency of the pre-clustering group identification, we introduce more notations and regularity conditions.
We denote the population distribution measure of responses as $P$, and denote the empirical measure as $P_n$.
For any probability measure $Q$ on $\mathbb{R}$ 
and any subset $\tilde{B}$ of $\mathbb{R}^{pL \times 1}$,
we define  $\Phi(\tilde{B},Q, \alpha) = \int \min_{
\tilde{b}\in \tilde{B} } \|Y- \alpha - \sum_{j=1}^p \gamma_j^T F_j^T \tilde{b}\|^2 Q(dY) $,
where $\gamma_j = \int_0^T X_j(t)\bm{B}(t)dt $ and
$F_j$ is defined similarly to $E_j$ in Equation (\ref{def_E_i}) with $n$ replaced by $p$.
When the responses follow a Gaussian distribution and ${\beta}_{(k),j}(t) \in S_{dM}$ for $k\in\{1,\dots,K\}, j\in\{1,\dots,p\}$, the empirical loss function in Equation (\ref{loss function: pre-clustering}) for the pre-clustering 
is equal to $\Phi(\tilde{B}(K), P_n, \alpha)$, where $\tilde{B}(K)$
consists of B-spline coefficients of $\beta_{(1)},\dots,\beta_{(K)}$.

We let $m_K(Q) = \inf\{\Phi(\tilde{B},Q, \alpha): \tilde{B} \text{ contains K or fewer points}, \alpha\in\mathbb{R}  \} $ denote the minimum of the loss function $\Phi$.
For a given $K$ and the sample size $n$, let 
$\tilde{B}_n = \tilde{B}_n(K) = \{\tilde{b}_{n(1)}, \dots, \tilde{b}_{n(K)}\}$ and $\tilde{\alpha}_n$ be the minimizer of $\Phi$ for $P_n$ such that $\Phi(\tilde{B}_n, P_n, \tilde{\alpha}_n) = m_K(P_n)$, 
where $\tilde{b}_{n(k)} = (\tilde{b}_{n,(k),1}^T, \dots, \tilde{b}_{n,(k),p}^T)^T$ for $k \in \{1, \dots, K\}$. 
Note that $\tilde{\alpha}_n $ and $\tilde{\beta}_{n(k)} = (\tilde{b}_{n,(k),1}^T \bm{B}(t),$ $\dots, \tilde{b}_{n,(k),p}^T \bm{B}(t))^T$ equal $ \tilde{\alpha}$ and $\tilde{\beta}_{(k)}$, respectively,  in Equation (\ref{minimizer: pre-clustering}) 
when we have Gaussian outcomes.
We also let 
$\Bar{B} = \Bar{B}(K) = \{ \Bar{b}_{(1)}, \dots, \Bar{b}_{(K)}\}$ and $\bar{\alpha}$ be the oracle versions of $\tilde{B}_n$ and $\tilde{\alpha}_n$, respectively, such that $\Phi(\Bar{B}, P, \Bar{\alpha}) = m_K(P)$, where $\Bar{b}_{(k)} = \{\Bar{b}_{(k),1}, \dots, \Bar{b}_{(k),p}\}$ for $k \in\{ 1, \dots, K\}$. 
The following are two additional regularity conditions.

(C8) For any $j\in \{1,\dots,K\},$ there exists unique $ \Bar{B}(j)$ and $\Bar{\alpha}$ \text{such that} $\Phi(\Bar{B}(j), P, \Bar{\alpha}) = m_j(P) $.


(C9) For any $j\in \{1,\dots,p\}$ and $l \in\{1,\dots,\mathcal{K}_j\} $, there exists a unique sub-sequence 
$\mathcal{A}_l \subset \{1,\dots, K \}$ and a constant $\tau>0$
such that for any
$k \in \mathcal{A}_l$ and $i\in \mathcal{G}_{j,l}  $,
we have $\| \left(\Bar{\beta}_{(k),j}(t) - {\beta}_{i,j}(t)\right)^2\|_\infty = O(1/n^\tau)$,  where  $\Bar{\beta}_{(k),j}(t) = \Bar{b}_{(k),j}^T \mathbf{B}(t)$.

Conditions (C8)  is analogous to assumptions in \citep{pollard1981strong}. 
Condition (C9) 
requires 
the oracle minimizer $\bar{\beta}_{(k),j} (t)$'s in the pre-clustering procedure to be close to  true coefficient functions.
Specifically,
for any true coefficient function $\beta_{ij} (t)$, there are some $\bar{\beta}_{(k),j} (t)$'s close to $\beta_{ij} (t)$ in terms of the infinity norm.

\begin{theorem}
\label{thm: pre-clustering}
When Conditions  (C2), (C5), (C6), (C7), (C8), and (C9) are satisfied and responses are Gaussian distributed,
we have
$ \Phi(\tilde{B}_n, P_n , \tilde{\alpha}_n) \xrightarrow{a.s.} m_K(P)$, and with probability one, the corresponding pre-clustering group identification $\tilde{\mathcal{G} }= \{ \tilde{\mathcal{G}}_{1},\dots, \tilde{\mathcal{G}}_{K} \}$ satisfies that  $\tilde{\mathcal{S}}(i) \neq \tilde{\mathcal{S}}(i')$ if $\mathcal{S}_j(i) \neq \mathcal{S}_j(i')$  for $i,i'\in \{1,2,\dots,n\}$, $j\in \{1,\dots,p \}$, where  $\tilde{\mathcal{S}}(i)$ denotes the pre-clustering group index for the $i$-th subject.
\end{theorem}




The almost sure convergence of 
$\Phi(\tilde{B}_n, P_n, \tilde{\alpha}_n)$ in Theorem \ref{thm: pre-clustering} implies that $\tilde{B}_n \xrightarrow{a.s.} \Bar{B}$ under Condition (C8). 
This  indicates that, with probability one, the empirical estimator $\tilde{\beta}_{(1)}, \dots, \tilde{\beta}_{(K)}$ defined in Equation \eqref{minimizer: pre-clustering} converge to the oracle minimizer $\Bar{\beta}_{(1)}, \dots, \Bar{\beta}_{(K)}$ 
 since $\Phi(\tilde{B}, P_n, \alpha)$ is equivalent to the pre-clustering loss function for Gaussian responses, where $\Bar{\beta}_{(k)}=(\Bar{\beta}_{(k),1}(t),\dots,\Bar{\beta}_{(k),p}(t))^T$.
Theorem \ref{thm: pre-clustering} also demonstrates that, with probability one, if two subjects belong to different true subgroups, they will not
be estimated into the same pre-clustering group.
In other words, the proposed pre-clustering groups provide a finer partition of the true subgroup identification, and it will not break the true subgroup structure.
}

{\color{black}
\section{Simulation}
\label{section: simulation}

In this section, we provide simulation studies to evaluate the performance of our method in comparison with existing methods including smooth functional linear model (SFLM) \citep{cardot2003spline}, recurrent neural network (RNN) \citep{rumelhart1986learning}, (generalized) linear regression models (LM/GLM), and response-based clustering (Resp) where data are first clustered by responses and then SFLM is applied in each cluster \citep{sun2022subgroup}. 
We use the relative root prediction mean squared error (RPMSE) and outcome misclassificarion rate (oMR) to evaluate prediction accuracy for continuous outcomes and binary outcomes, respectively, where $\text{RPMSE} = \text{RMSE}(y - \hat{y}) / \text{RMSE}(y)$, $\text{RMSE}(y - \hat{y}) = \left[\sum_{i=1}^n (y_i - \hat{y}_i)^2/n\right]^{1/2}$ and $\text{RMSE}(y) = \left[\sum_{i=1}^n y_i^2/n\right]^{1/2}$.
For estimation accuracy of estimators, we calculate a integrated squared error (ISE): $\text{ISE}(\hat{\beta}) = \| \bm{\hat{\beta}}(t) - \bm{\beta}(t)  \|_2/{\| \bm{\beta}(t)  \|_2}$, where \( \bm{\hat{\beta}}(t) \) is an estimator of  \( \bm{\beta}(t) \) and the norm  $\| \bm{\beta}(t) \|_2 = \left\{\int_0^T \beta_1(t)^2 \, dt  + \dots + \int_0^T \beta_n(t)^2 \, dt\right\}^{1/2}$.
We also calculate the subgroup misclassification rate (sMR) for the proposed method to assess the accuracy of subgroup identification. 


We generate continuous outcomes from the Gaussian model in Equation (\ref{model: HFGM}), and generate binary outcomes from the logistic model in Equation (\ref{model: HFLM}). 
To simplify illustration and mimic the real data in Section \ref{section: application}, we mainly focus on scenarios involving a single covariate $X_{i1}(t) =\bm{v}^T\bm{B_x}(t)$ for $i=1,2,\dots,n$, where  $\bm{v} \sim N_{26}(\bm{3},I_{26})$ and $\bm{B_x}(t)$ is the set of B-spline basis functions 
on time interval $[0, 23]$.
We provide simulation results with multiple covariates in the  Supplementary Materials, which are similar to those with only one single covariate, demonstrating that our method  performs  better than existing methods.


We consider the following three settings, where the first two contain continuous outcomes while the last one involves binary outcomes.

\begin{enumerate}[label=\textbf{\small Setting \arabic*}, align = left]
\item
\label{setting1}
We generate continuous responses from Equation (\ref{model: HFGM}), and let the number of true subgroups $\mathcal{K} = 4$, sample size $n = 100$ or $10000$, and coefficient functions  
$\beta_{i1}(t) = \bm{b_{i1}}^T\bm{B}(t)$, where
$\bm{b}_{i1} = 20 \mathbbm{I}( 1 \leq i \leq {n}/{4} ) \mathbf{1} + 
6 \mathbbm{I}( {n}/{4} + 1 \leq i \leq {n}/{2} ) \mathbf{1} -10\mathbbm{I}({n}/{2} + 1 \leq i \leq {3n}/{4} )  \mathbf{1} -40 \mathbbm{I}({3n}/{4} + 1 \leq i \leq n ) \mathbf{1} + \bm{\epsilon}'
$,
$\mathbbm{I}(\cdot)$ denotes the indicator function,
$\bm{1}_{35}$ represents a $35$-dimensional vector where all elements are equal to 1,
$ \bm{\epsilon}' \sim N_{35}(\bm{0},\sigma_{\epsilon'}^2 I_{35})$, and $\sigma_{\epsilon'} = 0.1, 0.5, 1, 5$, or $ 10$.

\item
\label{setting2}
Responses are also generated from Equation (\ref{model: HFGM}). We let the number of true subgroups $\mathcal{K} = 2$, sample size $n = 100$, and coefficient functions
$\beta_{i1}(t) = \sin(t)\mathbbm{I} ( 1 \leq i \leq {n}/{2} )  + \cos(t)\mathbbm{I} ( {n}/{2} + 1 \leq i \leq n ) $.

\item
\label{setting3}
We generate binary responses from the logistic model in Equation (\ref{model: HFLM}). We let the number of subgroups $\mathcal{K} = 2$, sample size $n = 100, 1000$ or $ 10000$, and $\beta_{i1}(t) = \bm{b_{i1}}^T\bm{B}(t)$ where
$\bm{b}_{i1} = 3 \mathbbm{I} ( 1 \leq i \leq {n}/{2} )  \mathbf{1} - 3 \mathbbm{I} ( {n}/{2} + 1 \leq i \leq n )  \mathbf{1} + \bm{\epsilon}',
$
$ \bm{\epsilon}' \sim N_{35}(\bm{0},\sigma_{\epsilon'}^2 I_{35})$, and $\sigma_{\epsilon'} = 1, 5$, or $10$.
\end{enumerate}


\begin{table}[h!]
\footnotesize
\centering
\begin{tabular}{c c c c c c c c c}
$\sigma_{\epsilon'}$ & Prop K=0 & Prop K=50 & Prop K=100 & Prop K=200 & SFLM & Resp & LM & RNN \\
\hline
\multicolumn{9}{c}{Case I: $n=100$} \\
\hline
$ 0.1$ & 0.0611 & --- & --- & --- & 1.0443 & 0.2504 & 1.0896 & 0.1614 \\
$ 0.5$ & 0.0630 & --- & --- & --- & 1.0399 & 0.2558 & 1.0894 & 0.1648 \\
$ 1$ & 0.0705 & --- & --- & --- & 1.0377 & 0.2567 & 1.0886 & 0.1685 \\
$ 5$ & 0.0735 & --- & --- & --- & 1.0363 & 0.2609 & 1.0874 & 0.1275 \\
$ 10$ & 0.0748 & --- & --- & --- & 1.0440 & 0.2670 & 1.0857 & 0.1475 \\
\hline
\multicolumn{9}{c}{Case II: $n=10000$} \\
\hline
$ 0.1$ & --- & 0.0038 & 0.0036 & 0.0036 & 0.8371 & 0.2261 & 0.9884 & 0.0186 \\
$ 0.5$ & --- & 0.0040 & 0.0037 & 0.0037 & 0.8418 & 0.2291 & 0.9592 & 0.0198 \\
$ 1$ & --- & 0.0046 & 0.0044 & 0.0044 & 0.8525 & 0.2389 & 0.9884 & 0.0186 \\
$ 5$ & --- & 0.0049 & 0.0050 & 0.0047 & 0.8932 & 0.2418 & 0.9714 & 0.0210 \\
$ 10$ & --- & 0.0049 & 0.0050 & 0.0049 & 0.9029 & 0.2472 & 0.9884 & 0.0186 \\
\hline
\end{tabular}
\caption{Relative root prediction mean squared error (RPMSE) of all methods under \ref{setting1} with varying values of $\sigma_{\epsilon'}$.``Prop K=50/100/200" represents the proposed method with $K$ pre-clustering groups. For $n=100$, no pre-clustering is applied, and the RPMSE is listed under ``Prop K=0." 
SFLM, Resp, LM, and RNN  stand for the smooth functional linear model, the response-based clustering method, the linear regression  model, and the recurrent neural network, respectively.
}
\label{table: 4 group RPMSE}
\end{table}

\begin{table}[h!]
\footnotesize
\centering
\begin{tabular}{c c c c c c c c}
$\sigma_{\epsilon'}$ & Prop K=0 & Prop K=50 & Prop K=100 & Prop K=200  & SFLM & Resp & {sMR} \\ 
\hline
\multicolumn{8}{c}{Case I: $n=100$} \\
\hline
$ 0.1$ & 0.61 & --- & --- & --- & 4.12  & 1.77 & 6.92\% \\
$ 0.5$ & 0.67 & --- & --- & --- & 4.09 & 1.79 & 7.08\% \\
$ 1$   & 0.73 & --- & --- & --- & 4.17 & 1.79 & 7.72\% \\
$ 5$   & 0.85 & --- & --- & --- & 4.11 & 1.82 & 9.21\% \\
$ 10$  & 0.98 & --- & --- & --- & 4.18 & 1.83 & 9.42\% \\
\hline
\multicolumn{8}{c}{Case II: $n=10000$} \\
\hline
$ 0.1$ & --- & 0.070 & 0.072 & 0.069 & 3.19 & 0.30 & 4.59\% \\
$ 0.5$ & --- & 0.071 & 0.073 & 0.077 & 3.21 & 0.31 & 4.48\% \\
$ 1$   & --- & 0.083 & 0.082 & 0.082 & 3.28 & 0.33 & 4.59\% \\
$ 5$   & --- & 0.095 & 0.098 & 0.094 & 3.18 & 0.33 & 4.66\% \\
$ 10$  & --- & 0.110 & 0.108 & 0.099 & 3.14 & 0.35 & 4.59\% \\
\hline
\end{tabular}
\caption{ Integrated squared error (ISE) of different methods 
and subgroup misclassification rate (sMR) of the proposed method under \ref{setting1} with different $\sigma_{\epsilon'}$'s. The first five columns contain ISE values, and the last column provides sMR of the proposed method. RNN and LM are excluded as they do not produce functional estimators.
}
\label{table: 4 group ISE MR}
\end{table}

In \ref{setting1},  we investigate the performance of all the methods when distinction between different subgroups varies, 
where the distinction increases as $\sigma_{\epsilon'}$ decreases. 
As shown in Tables \ref{table: 4 group RPMSE} and \ref{table: 4 group ISE MR},
for each $\sigma_{\epsilon'}$,
the proposed method outperforms existing methods.
For instance, when $n=100$ and $\sigma_{\epsilon'} = 1$,  the proposed method achieves a 58.16\% improvement in RPMSE compared to RNN, and a 59.22\% improvement in ISE compared to Resp. 
We can also observe that sMR in all cases is low, indicating that our method can recover the true subgroup structure.
Moreover, the proposed method and the Resp perform better when $\sigma_{\epsilon'}$ is smaller, while  changes of other methods are slight.
In particular, when the sample size $n$ is fixed,  RPMSE, ISE, and sMR of the proposed method decrease as $\sigma_{\epsilon'}$ decreases. 

For cases with a large sample size $n=10000$, 
we apply the proposed pre-clustering procedure with $K=50, 100,$ or $200$ in the proposed method.
Based on the results in Tables \ref{table: 4 group RPMSE} and \ref{table: 4 group ISE MR}, we can see that the choice of $K$ (the number of pre-clustering groups) does not substantially affect the performance of our method, provided that it is much larger than the true number of subgroups. This offers flexibility in the selection of $K$.
Also, the efficacy of the proposed method increases in terms of RPMSE and ISE when we have a larger sample size. 

\begin{table}[h]
\small
\centering
\begin{tabular}{cccccccccc}
$L$ & {Prop} & {SFLM} & {Resp} & {LM} & {RNN} & {ISE Prop} & {ISE SFLM} & {ISE Resp} & {sMR} \\ 
\hline
$ 20  $ & 0.31 & 0.96 & 0.75 & 0.84 & 0.33 & 2.14 & 2.71 & 2.49 & 15.23\% \\
$ 30  $ & 0.28 & 0.98 & 0.51 & 0.84 & 0.33 & 2.12 & 2.83 & 2.25 & 13.37\% \\
$ 40  $ & 0.26 & 1.05 & 0.51 & 0.84 & 0.33 & 2.12 & 3.13 & 2.24 & 13.19\% \\
$ 50  $ & 0.25 & 1.09 & 0.49 & 0.84 & 0.33 & 2.09 & 3.21 & 2.22 & 12.43\% \\
\hline
\end{tabular}
\caption{
Relative root prediction mean squared error (RPMSE),
Integrated squared error (ISE) for different methods, 
and subgroup misclassification
rate (sMR) for the proposed method under \ref{setting2} with different values of $L$ (the number of basis functions). 
The first five columns present the RPMSE values for all the methods, while columns 6 through 8 showcase the ISE values for each method.
SFLM, Resp, LM, and RNN  stand for the smooth functional linear model, the response-based clustering method, the linear regression  model, and the recurrent neural network, respectively.
}
\label{tab:sim_results_continuous_beta}
\end{table}

In \ref{setting2}, 
the distinction between coefficient functions of different subgroups is less clear than that in \ref{setting1}.
The results are provided in Table \ref{tab:sim_results_continuous_beta}.
Although the improvement of our proposed method under \ref{setting2} is not  as much as that under \ref{setting1}, our approach still outperforms existing methods. 
For instance, when we use $L=40$ (the number of basis functions), the proposed method can achieve a 21.2\% improvement in RPMSE compared to RNN, and a 5.86\% improvement in ISE compared to Resp.
The subgroup misclassification rate is 13.19\%, indicating that the majority of subjects are correctly classified.

{
Moreover, we investigate the impact of the choices of the number of basis functions, $L$, in the proposed method under \ref{setting2}. Results in Table \ref{tab:sim_results_continuous_beta} indicate that the proposed method is robust when $L$ is sufficiently large. Although more basis functions improve accuracy, they also increase computational burden. The results suggest that a moderate number of basis functions, such as 40, balances performance and computational efficiency of the proposed method.

}

\begin{table}[H]
\centering
\resizebox{\textwidth}{!}{%
\begin{tabular}{cccccccccc}
$\sigma_{\epsilon'}$ & oMR Prop & oMR SFLM & oMR Resp & oMR GLM & oMR RNN & ISE Prop & ISE SFLM & ISE Resp & {sMR}\\ 
\hline
\multicolumn{10}{c}{Case I: $n=100$} \\
\hline
$ 1  $ & 9.18\%  & 18.35\% & 13.98\% & 27.12\% & 11.23\% & 0.94  & 11.39 & 2.54 & 6.38\% \\
$ 5  $ & 15.29\% & 24.11\% & 19.44\% & 33.05\% & 17.14\% & 1.04  & 13.22 & 2.59 & 6.76\% \\
$ 10 $ & 18.43\% & 27.07\% & 22.57\% & 39.18\% & 19.21\% & 1.33 & 19.28 & 2.61 & 7.12\% \\
\hline
\multicolumn{10}{c}{Case II: $n=1000$} \\
\hline
$ 1  $ & 7.14\%  & 17.18\% & 12.76\% & 24.23\% & 10.12\% & 0.83  & 10.84 & 2.44 & 5.34\% \\
$ 5  $ & 12.28\% & 20.30\% & 18.38\% & 30.17\% & 15.25\% & 0.99  & 13.18 & 2.48 & 5.48\%\\
$ 10 $ & 16.32\% & 24.24\% & 20.60\% & 34.31\% & 18.19\% & 1.00  & 17.30 & 2.49 & 5.77\%\\
\hline
\multicolumn{10}{c}{Case III: $n=10000$} \\
\hline
$ 1  $ & 6.21\%  & 16.12\% & 10.94\% & 22.07\% & 8.08\% & 0.73  & 10.18 & 2.18 & 3.30\%\\
$ 5  $ & 9.37\%  & 17.27\% & 15.81\% & 27.13\% & 11.14\% & 0.60  & 12.49 & 2.19 & 3.75\%\\
$ 10 $ & 12.41\% & 21.21\% & 17.63\% & 33.29\% & 13.92\% & 0.53  & 15.24 & 2.22 & 4.14\% \\
\hline
\end{tabular}
}
\caption{
Outcome misclassification rate (oMR),
integrated squared error (ISE),
and subgroup misclassification rate  (sMR) for different methods under  \ref{setting3} with varying values of $\sigma_{\epsilon'}$.
In the proposed method, a pre-clustering procedure with $K=100$ is applied when $n=1000$ or $n=10000$.
SFLM, Resp, GLM, and RNN stand for the smooth functional linear model, the response-based clustering method, the generalized linear model, and the recurrent neural network, respectively.
}
\label{tab:binary_simresults}
\end{table}

In \ref{setting3}, outcomes are binary and thus we use oMR instead of RPMSE as a criterion for predictive performance. 
Results in Table \ref{tab:binary_simresults} suggest that the oMR and ISE of the proposed method are smaller than those of other methods. 
For instance, when $n=1000$ and $\sigma_{\epsilon'} = 5$, the proposed method can achieve a 19.48\% improvement in oMR compared to RNN, and a 60.08\% improvement in ISE compared to Resp.
To further illustrate this, we provide a corresponding figure {
in the  Supplementary Materials} depicting estimation results of coefficient functions by the proposed method in a specific simulation replication when $n=100, \sigma_{\epsilon'} = 1$. The estimation results from the proposed method reveal two subgroups, which is consistent with the correct number of true subgroups. 
Remarkably, the figure suggests that the estimated coefficient functions and true coefficient functions are highly close, which indicate that our method possesses effective performance.



\section{Real Data Application}
\label{section: application}

In this section, we apply the proposed method and existing methods to physical activity and diseases data in  UKB.  
We examine two types of diseases: Parkinson's disease  and mental disorders.
For Parkinson's disease, the outcome is binary with `1' and `0' corresponding to 
subjects with and without, respectively, Parkinson's disease diagnoses.
There are 80,692 subjects with available physical activity and Parkinson's diagnosis, of whom 438 have been diagnosed with Parkinson's disease.
Mental disorders are assessed through the neuroticism score which consists of 13 ordinal levels.
Our dataset includes 79,252 individuals with valid data for both physical activity and neuroticism scores.
Similarly to simulations, we compare the performance of the proposed method and existing methods: SFLM, LM/GLM and Resp. Here we do not include the RNN since our proposed method outperforms it in simulation studies and the computation cost of RNN is high for large sample sizes. 
Moreover, RNN cannot estimate subgroup identification and lacks the interpretability needed to provide scientific insights into the relationship between physical activity and diseases. 

To assess the predictive efficiency of methods, we utilize physical activity data 
from the first entire day 
as the training set, while  the subsequent four days as the testing set. 
We utilize predictive false negative rate (PFNR), predictive false positive rate (PFPR), and predictive area under curve (PAUC) to evaluate predictive accuracy for Parkinson's disease. 
PFNR and PFPR are defined as the average values of false negative rate (FNR) and false positive rate (FPR) across the four testing days, respectively, and PAUC  stands for the average area under receiver operating characteristic (ROC) curves across four testing days.
For  neuroticism scores, we evaluate the predictive performance through  root predictive mean squared error (RPMSE), which is defined as the square root of the average mean squared error (MSE) across four testing days, expressed as
$
\text{RPMSE} = {1}/{4}\sum_{j=1}^4 \sqrt{{1}/{n}\sum_{i=1}^n (y_{ij} - \hat{y}_{ij} )^2}
$
where $\hat{y}_{ij}$ denotes the estimated neuroticism score of the \(i\)-th subject on the \(j\)-th testing day.

\subsection{Parkinson's Disease}
\label{subection: PD}

We apply the proposed pre-clustering and HFLM  to physical activity data and Parkinson's disease data and set the number of pre-clustering groups to be 100.  
The PFNRs, PFPRs and PAUCs of all methods are presented in Table \ref{table: Parkinson pred}. 
The proposed method reduces the PFNR of each existing method by at least 77\% while  maintaining a low PFPR.
Although the PFPR of our method is slightly higher than that of other methods, it remains very close to zero at a value of 0.002.
Additionally, the proposed method achieves a higher PAUC than existing methods. 
Specifically, the proposed method increases at least  29.49\%  of the PAUC of each existing method.

\begin{table}[h]
\small
\centering
\begin{tabular}{cccc}
Methods & Proposed & SFLM & GLM \\ 
\hline
PFNR & 0.210 & 0.923 & 1.000 \\ 
PFPR & 0.002 & 0.000 & 0.000 \\ 
PAUC & 0.966 & 0.691 & 0.712 \\ 
\hline
\end{tabular}
\caption{
Predictive performance of different methods for Parkinson's disease. PFNR, PFPR, and PAUC represent the predictive false negative rate, the predictive false positive rate, and  the predictive area under the curve, respectively. SFLM, Resp, and GLM stand for the smooth functional linear model, the response-based clustering method, and the generalized linear model, respectively.
}
\label{table: Parkinson pred}
\end{table}

{
The high PFNRs of traditional methods may be due to the 
imbalanced Parkinson's disease outcomes,
where only 5\% of all subjects are diagnosed with this disease.
By accounting for heterogeneity, our method identifies a more balanced  subgroup with over 17\% of individuals diagnosed with the Parkinson's disease,
facilitating the detection of subjects with the disease and resulting in a lower PFNR.
Besides the 17\% diagnosed cases, 
the remaining participants in this subgroup may have a higher risk of Parkinson's disease than those in other subgroups.
In fact, individuals diagnosed with Parkinson's disease or at high risk of developing Parkinson's disease may share inherent characteristics that differ from those of the general population \citep{fang2018association, schalkamp2023wearable}, which may lead to heterogeneous effects 
across subjects.
}
}

{\color{black}
Our proposed method identifies three distinct subgroups.
Figure \ref{fig: Parkinson subgroup plots} provides  estimated coefficient functions of the subgroups, with the time axis spanning a full day from 00:00 to 24:00. 
All the coefficient functions
are negative during  the majority of time,
which aligns with existing scientific discovery that physical activity tends to mitigate the risk of Parkinson's disease \citep{chen2005physical, speelman2011might}.

\begin{figure}[h]
\centering  
\includegraphics[width=0.9\textwidth]{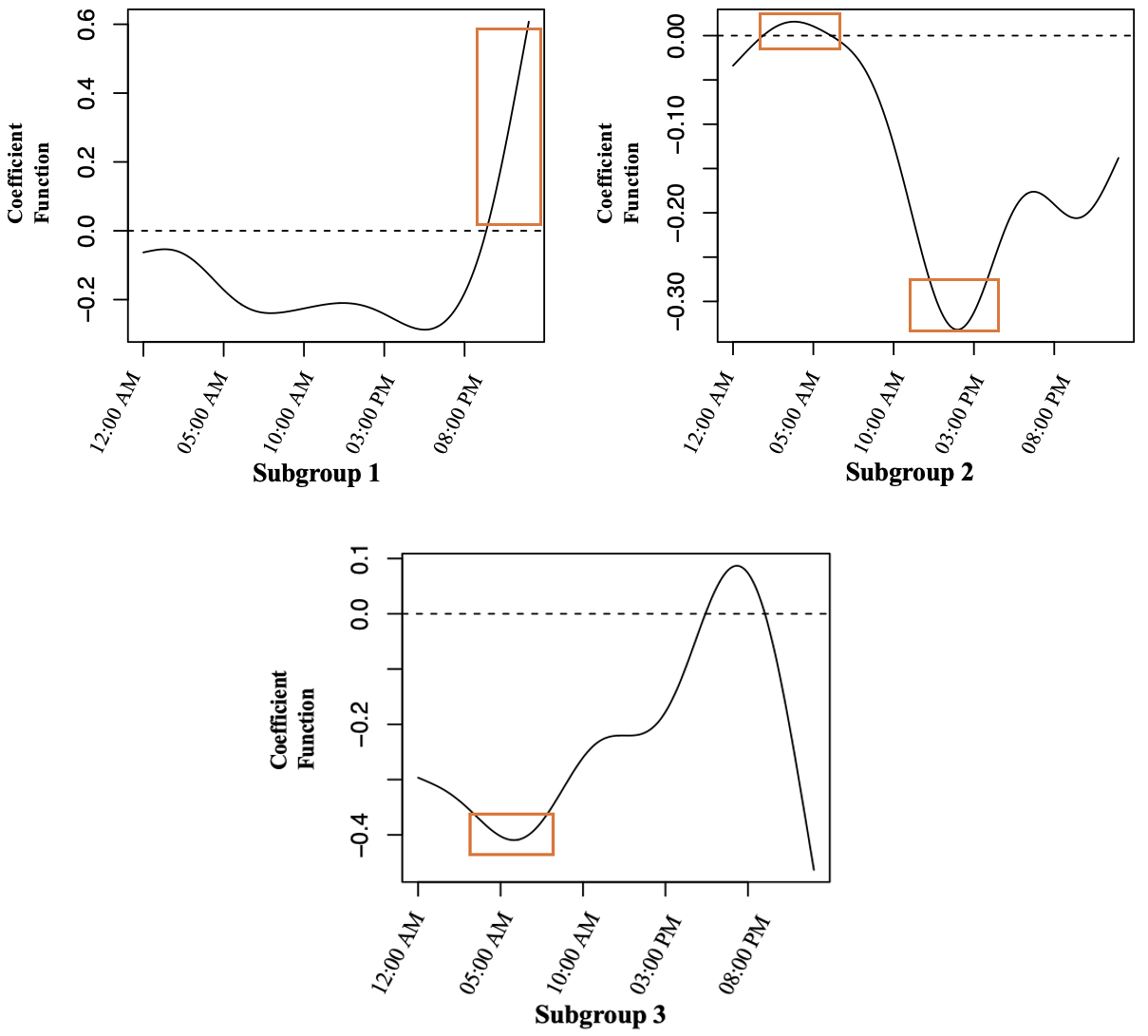}
\vspace{10pt}
\caption{ Estimated coefficient functions of the relationship between physical activity and Parkinson's disease by the proposed method.
The x-axis represents real time in hours.
The part highlighted by the orange box is analyzed and explained in Subsection \ref{subection: PD}}.
\label{fig: Parkinson subgroup plots}
\vspace{10pt}
\end{figure}

Subgroup 1 
is the subgroup 
with over $17\%$ of the subjects diagnosed with Parkinson's disease,
which is much more than that of other subgroups.
As shown in Figure \ref{fig: Parkinson subgroup plots},
the coefficient function of this subgroup exhibits a distinct pattern compared to other subgroups.
The coefficient function of subgroup 1 is positive between 9:00 PM and 12:00 AM with a relatively large magnitude, which indicates a positive relationship between nighttime physical activity and Parkinson's disease risk.
This positive relationship is possibly due to that individuals diagnosed with or at high risk for Parkinson's disease often have sleep problems, leading to frequent nighttime awakenings and, consequently, increased physical activity during the night \citep{kumar2002sleep, dhawan2006sleep}.
We also observe that the coefficient function for subgroup 1 is more negative from 8:00 AM to 6:00 PM than the remaining time within a day, suggesting that physical activity during this time period may reduce the risk of Parkinson's disease for individuals in this subgroup more than other time.

The coefficient function for subgroup 2 is positive between 2:00 AM and 6:00 AM, suggesting that early morning waking or nocturnal awakenings may be positively associated with an increased risk of Parkinson's disease, which is consistent with existing literature \citep{bhidayasiri2018getting}.
For subgroup 2, the most negative part of the coefficient function occurs around 2:30 PM, indicating that exercising in the early afternoon, particularly after lunch, may help reduce Parkinson's disease risk for individuals in this group.
In contrast, for subgroup 3, the coefficient function during the daytime is the most negative at around 7:00 AM,
suggesting that individuals in subgroup 3 may benefit more from exercising early in the morning. 
This comparison highlights that optimal exercise times may vary across the two subgroups, 
where afternoon exercise is more effective for subgroup 2, while
morning exercise is more effective for subgroup 3.
}




{\color{black}
\subsection{Neuroticism Score}
\label{subsection: neuroticism}
We apply the proposed pre-clustering and HFGM to physical activity data and neuroticism scores  in the UKB and  
set the number of pre-clustering  groups to be 100. 
The RPMSEs of all methods are presented in Table \ref{table: neuro}, which shows that proposed method achieves a much lower RPMSE compared to existing methods. Specifically, 
for each existing method, the proposed method reduces at least 20\% of its corresponding RPMSE.

\begin{table}[h]
\small
\centering
\begin{tabular}{cccc}
Methods & Proposed & SFLM & LM \\ 
\hline
RPMSE & 1.62 & 3.42 & 3.17 \\ 
Percentage of Reduction in RPMSE &  & 52.63\% & 48.89\% \\ 
\hline
\end{tabular}
\caption{
Root predictive mean squared errors (RPMSEs) for different methods on neuroticism scores.
SFLM, Resp, and LM  represent the smooth functional linear model, the response-based clustering method, and the linear regression model, respectively.
}
\label{table: neuro}
\end{table}

\begin{figure}[h]
\label{fig: neuro}
\centering  
\includegraphics[width=1\textwidth]{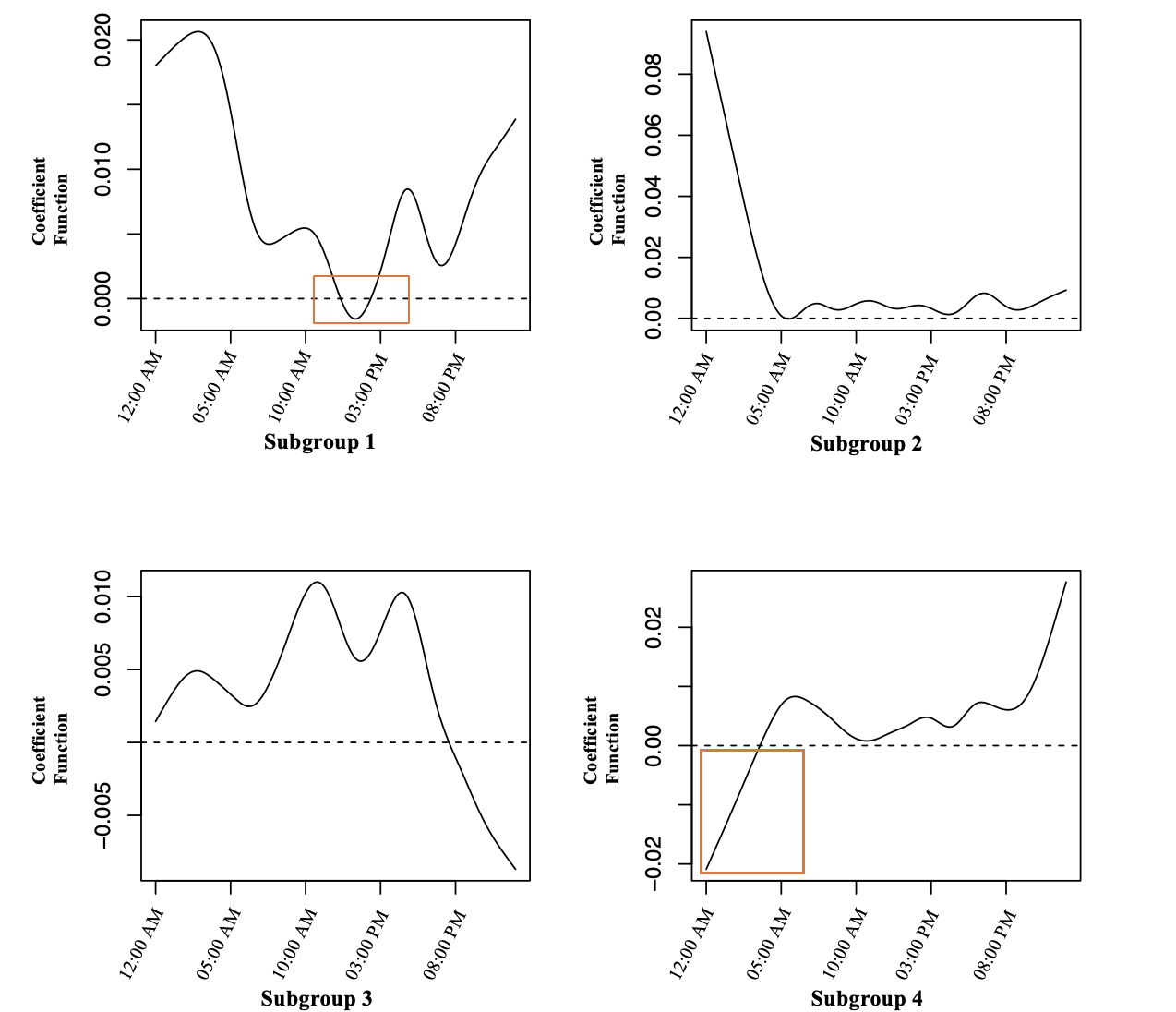}
\caption{Estimated coefficient functions of the relationship between physical activity and neuroticism score by the proposed method.
The x-axis represents real time in hours.
The part highlighted by the orange box is analyzed and explained in Subsection \ref{subsection: neuroticism}.
}
\label{fig: neuro subgroup coef plots}
\end{figure}

Our proposed method identifies four distinct subgroups with estimated coefficient functions provided in
Figure \ref{fig: neuro subgroup coef plots}.
For the first three subgroups, the coefficient functions are positive from 12:00 AM to 5:00 AM, indicating a positive relationship between physical activity and neuroticism score during this time period. 
This suggests that excessive nighttime activity, such as that caused by insomnia, can increase the risk of mental disorders for most individuals, consistent with findings from medical research \citep{beltagy2018night, tanaka2004sleep, tancredi2022artificial}.
In contrast, for the last subgroup, the coefficient function is negative during this time period, indicating that nocturnal activity may reduce the risk of mental disorders for these individuals \citep{facer2019resetting}. This could be attributed to genetic factors or lifestyle habits, as some people are naturally ``night owls''. For individuals like novelists or artists, who often feel more creative and productive at night, engaging in midnight activity may help them feel more at ease, thus lowering their risk of mental disorders \citep{kanazawa2009night}.

Distinct patterns also emerge within the first three subgroups.
For subgroup 2, the magnitude of the coefficient function is large from 12:00 AM to
5:00 AM, compared to that 
during
the rest of the day.
This suggests that the mental health of the individuals in this subgroup is particularly sensitive to nighttime activity \citep{rebar2014differences}. 
As a result, maintaining good sleep is crucial for this group.
For subgroup 1, the coefficient function is positive at the end of the day and negative between 12:30 PM and 2:30 PM. This suggests that reducing nighttime physical activity before sleep may be beneficial for people in this subgroup, while the most effective time for anxiety-reducing exercise is after lunch, aligning with discoveries in \cite{mackay2010effect}.
Although the magnitude of the coefficient function for subgroup 1 is small after lunch, the effect of physical activity is also influenced by the magnitude of the activity, which is much higher during the day than at night. 
For the third subgroup, the coefficient function is negative from 8:00 PM to 12:00 AM, implying that people in this subgroup are suggested to engage in activities or exercise after dinner to reduce the risk of mental disorder \citep{ji2022physical}.

We also conduct one-way ANOVAs  and  Chi-square tests to investigate whether there is a difference across the four subgroups in demographic and socioeconomic characteristics: age, 
gender, and job type. 
The results are provided in the Supplementary Materials.

}

\section{Discussion}
\label{section: discussion and conclusion}
Our main contribution is to enhance the understanding of the potential heterogeneous functional effect of physical activity on diseases.
Specifically, we propose a GHFM within the generalized
FDA framework to capture this heterogeneous functional effect, and develop a pre-clustering method to improve computational efficiency for large-scale data.
The proposed methods are novel in their ability to handle large mobile health datasets and to identify subgroups with time-varying heterogeneous effects without predefining the number of subgroups, ensuring accurate insights and reduced computational cost.
It supports both continuous and generalized outcomes, making it versatile for real-world applications, and demonstrates superior future-day prediction accuracy over existing methods in large datasets such as UKB data.

{
In the future, we could develop inference methods to quantify the uncertainty of our estimators. Specifically, this could involve conducting hypothesis tests and confidence intervals for the identified subgroup structure and estimated coefficients.}
Moreover, exploring the nature of the subgroups identified by our method in more depth is also valuable. We may investigate factors associated with these subgroups, including potential correlations between genetic variables and subgroup classifications.
Finally, our methods could also be applied to other large-scale datasets, such as the ``All of Us" dataset \citep{all2019all}, for further discoveries.

\section*{Supplementary Materials}
 Algorithms, proofs, additional simulation results, and additional details on the real data application can be found in the supplementary materials. 

\section*{Acknowledgments}
We gratefully acknowledge support  from the UK Biobank study and the Purdue Rosen Center For Advanced Computing (RCAC). This work was also supported by the National Science Foundation under Grant DMS 2210860.

\section*{Disclosure Statement}
The authors report that there are no competing interests to declare.

\begingroup
\setstretch{1.1}  
\small
\footnotesize
\bibliographystyle{Chicago}
\bibliography{hetero_ref_upper_force.bib}
\endgroup


\end{document}